\documentclass{PoS}
\usepackage{graphicx}
\usepackage{amssymb,amsmath,latexsym,enumerate}

\title{On the mechanisms of the quasi-biennial\\
oscillations in the GCR intensity}

\ShortTitle{On the QBO in the GCR intensity}

\author{\speaker{M. Krainev}, G. Bazilevskaya, M. Kalinin, A. Svirzhevskaya, N. Svirzhevsky\\
Lebedev Physical Institute, Moscow, Russia\\
E-mail: \email{mkrainev46@mail.ru},\\
\email{gbaz@rambler.ru}, \email{mkalinin@fian.fiandns.mipt.ru},\\
\email{svirzhak@fian.fiandns.mipt.ru}, \email{svirzhev@fian.fiandns.mipt.ru}
}

\abstract{
Quasi-biennial oscillation (QBO) is a well-known quasi-periodical variation with characteristic time 0.5-4 years in different solar, heliospheric and cosmic ray characteristics. Recently it has been shown that there are low correlation between the solar and heliospheric QBOs and rather high anticorrelation between the QBOs in galactic cosmic ray (GCR) intensity near the Earth and in the strength of the heliospheric magnetic field (HMF). Besides, it was suggested that both step-like changes of the GCR intensity and Gnevyshev Gap effect (a temporal damping of the solar modulation around the sunspot maxima) could be viewed as the manifestations of QBO. Some suggestions were also made on the mechanisms of QBO in the GCR intensity.

In this paper a hypothesis is checked on the causes of the apparent lack of correlation between solar and heliospheric QBOs, then the possible mechanisms of QBO in the GCR intensity are discussed as well as the idea of the same nature of the step-like changes and Gnevyshev Gap effects in the GCR intensity.

{\underline{Our main conclusions}} are as follows:
\begin{enumerate}
\item
In the first approximation the hypothesis is justified that the change in the sunspot and QBO cycles in the transition from the Sun to the heliosphere is due to 1) the different magnitude and time behavior of the large-scale and small-scale photospheric solar magnetic fields and 2) the stronger attenuation of the  small-scale fields in this transition.
\item
As the QBO in the HMF strength influences both the diffusion coefficients and drift velocity, it can give rise to the complex QBO in the GCR intensity with respect to the dominating HMF polarity. The description of drift velocity field for the periods of the HMF inversion is suggested, although it has drawbacks.
\item
As the conditions in the heliosphere are quite different around the sunspot maximum and during the periods of low solar activity (both with respect to the HMF polarity distribution and with the presence or absence of the large-scale barriers),
the suggestion that both the step-like changes of the GCR intensity and Gnevyshev Gap effect could have the same nature, looks questionable.
\end{enumerate}
}
\FullConference{The 34th International Cosmic Ray Conference,\\
		30 July- 6 August, 2015\\
		The Hague, The Netherlands}
\begin{document}

\section{Introduction}
\noindent Quasi-biennial oscillation (QBO) is a well-known quasi-periodical variation with characteristic time 0.5-4 years in different solar, heliospheric and cosmic ray characteristics, see \cite{Bazilevskaya_etal_SSR_186_359_2014,Bazilevskaya_etal_JoP_CS_inprint_2015,Bazilevskaya_etal_CosRes_inprint_2016} and references therein. QBO appears to be the most prevalent quasi periodicity shorter than the 11-year cycle in solar activity phenomena.
Here we are interested, mainly, in three aspects among the many facts summarized in \cite{Bazilevskaya_etal_SSR_186_359_2014,Bazilevskaya_etal_JoP_CS_inprint_2015,Bazilevskaya_etal_CosRes_inprint_2016}.
First, in \cite{Bazilevskaya_etal_SSR_186_359_2014} it has been shown that there is a lack of correlation between solar and heliospheric QBOs: the correlation coefficient between the QBO in the sunspot area $S_{ss}$, $QBO_{Sss}$, and the QBO in the regular heliospheric magnetic field (HMF) strength $B^{hmf}$, $QBO_{Bhmf}$, is $\rho\approx 0.2$. This correlation increases up to $\rho\approx 0.5$ if the heliospheric QBOs are shifted behind the solar ones by about 3 months.
Second, in \cite{Bazilevskaya_etal_SSR_186_359_2014} it was shown that the QBOs in the galactic cosmic ray intensity $J_{gcr}$, $QBO_J$, are not coherent
with $QBO_{Sss}$ and other solar indices, while they correspond well with $QBO_{Bhmf}$ and it was suggested that both step-like changes of the GCR intensity and Gnevyshev Gap (GG) effect (a temporal damping of the solar modulation around the sunspot maxima, see the references in \cite{Bazilevskaya_etal_SSR_186_359_2014}) could be viewed as the manifestations of $QBO_J$. Third, in \cite{Bazilevskaya_etal_JoP_CS_inprint_2015,Bazilevskaya_etal_CosRes_inprint_2016} it was suggested that small delay of the $QBO_J$ relative to $QBO_{Bhmf}$ argues for a diffusion mechanism of the $QBO_J$ acting within $\approx 10$ AU from the Sun , while the difference in the correlation coefficients for the periods with the dominating HMF polarity ($A$, the sign of $B^{hmf}_r$ in the N-hemisphere) $A>0$ and $A<0$ may be indicative of the drift influence.

In this paper, after introducing necessary definitions and describing the data used, we formulate and check a hypothesis on the causes of the  apparent lack of correlation between solar and heliospheric QBOs using as a proxy of both solar magnetic fields (SMF) and HMF the energy indices, introduced in \cite{Krainev_etal_Pulkovo_Conf_121_1999}. Then the possible mechanisms of  $QBO_J$ are considered in slightly more details and we critically discuss the idea of the same nature of the step-like changes of the GCR intensity during the periods of low solar activity and GG-effect during the solar maximum phases.

\section{Definitions and data}
\noindent
In this paper as a proxy for QBO in the time series of any characteristic $P$, monthly or Carrington rotation (CR) or Bartels rotation (BR) averaged, we consider the same as in \cite{Bazilevskaya_etal_SSR_186_359_2014,Bazilevskaya_etal_JoP_CS_inprint_2015,Bazilevskaya_etal_CosRes_inprint_2016} simple and robust expression $QBO_P=P_7-P_{25}$, where $P_k$ means the $P$-series smoothed with the period of $k$ points. As a proxy for a long-term or sunspot cycle (SC) in the same characteristic we use $SC_P=P_{13}$, that is, approximately yearly smoothed $P$-series. Besides the CR time series of the SMF energy indices (see the next section), calculated using the Wilcox Solar Observatory (WSO), USA, data and models \cite{WSO_Site}, we also use the CR series on the sunspot area $S_{ss}$ from \cite{Sss_Site} and the BR time series on the HMF strength near the Earth $B^{hmf}$ from \cite{OMNI_Site}.

\section{Transition of QBO from the Sun to the heliosphere}
\noindent In \cite{Bazilevskaya_etal_JoP_CS_inprint_2015,Bazilevskaya_etal_CosRes_inprint_2016} it was shown that the sunspot area and the photospheric SMF energy index, introduced in \cite{Krainev_etal_Pulkovo_Conf_121_1999}, change similar both in their long-term (SC) and QBO variations. Using the WSO data and models, the SMF energy indices could be constructed not only for photosphere, $r_{ph}$, but for any radial distance between $r_{ph}$ and the source surface $r_{ss}=(2.5\div 3.25)r_{ph}$, in the transition layer from the Sun to the heliosphere. In this layer the energy density of the magnetic fields dominates over the solar wind thermal and kinetic energy densities, fixing the main features of the solar wind and HMF in the heliosphere, see \cite{WSO_Site,Krainev_Webber_IAUSymp_223_81_2004}. Probably these indices could contain some valuable information on the space and time structure of the magnetic fields in this very important region. So before proceeding further with the discussion of QBOs it could be useful to say some words on the WSO models and the SMF energy indices.

In the most widely used potential-field-source-surface WSO model, the radial SMF component $B_r$ in the range $r_{ph}\le r\le r_{ss}$ is expressed in spherical coordinates $\{r,\vartheta,\varphi\}$ as a series in terms of
spherical functions $Y_{lm}(\vartheta,\varphi)$ of degree $l$ and order $m$:
\begin{align}
B_r(r,\vartheta,\varphi;r_{ss})&=\sum_{l=0}^9 \sum_{m=-l}^l a_{lm} C_r(r;l,r_{ss})
Y_{lm}(\vartheta,\varphi)=\sum_{l=0}^9 B_r^l(r,\vartheta,\varphi;r_{ss})
\label{Br}
\end{align}

The expressions for $B_\vartheta$ and $B_\varphi$ can be written in the
similar way,
$C_r$, $C_\vartheta$, $C_\varphi$ being the known functions. The complex
coefficients $a_{lm}$, or
rather, their real counterparts $g_{lm}$ and $h_{lm}$ are available on the
WSO home page \cite{WSO_Site} for both types of the inner boundary conditions, fixing from observations the line-of-sight photospheric SMF component $B_{ls}^{ph}$ (in the ``classic'' variant of the WSO model) or $B_r^{ph}$ (in the ``radial'' variant). In Eq. (\ref{Br}) we also represent $B_r$ as a sum of the partial $B_r^l$ due to the SMF with the same degree $l$.

In our works on the structure of the solar cycle maximum phase we used the SMF energy index introduced in \cite{Obridko_Shelting_SP_137_167_1992}. Generalizing, in \cite{Krainev_etal_Pulkovo_Conf_121_1999,Bazilevskaya_etal_SolPhys_197_157_2000,Bazilevskaya_etal_JoP_CS_inprint_2015,Bazilevskaya_etal_CosRes_inprint_2016} we discussed the behavior of the energy index of the radial SMF component
integrated over all longitudes and latitudes on the
photosphere, but calculated without the monopole term ($l=0$) in Eq. (\ref{Br}),
\begin{equation}
\verb|Br2_PH|=
\int_0^{\pi}\int_0^{2\pi}{B_r}^2(r_{ph},\vartheta,\varphi)\sin\vartheta d\vartheta d\varphi,
\label{Br2_PH}
\end{equation}
\noindent and also the similar energy index on the source surface, \verb|B2_SS|.

\begin{figure}[h]
  \centering
  \includegraphics[width=1.\textwidth]{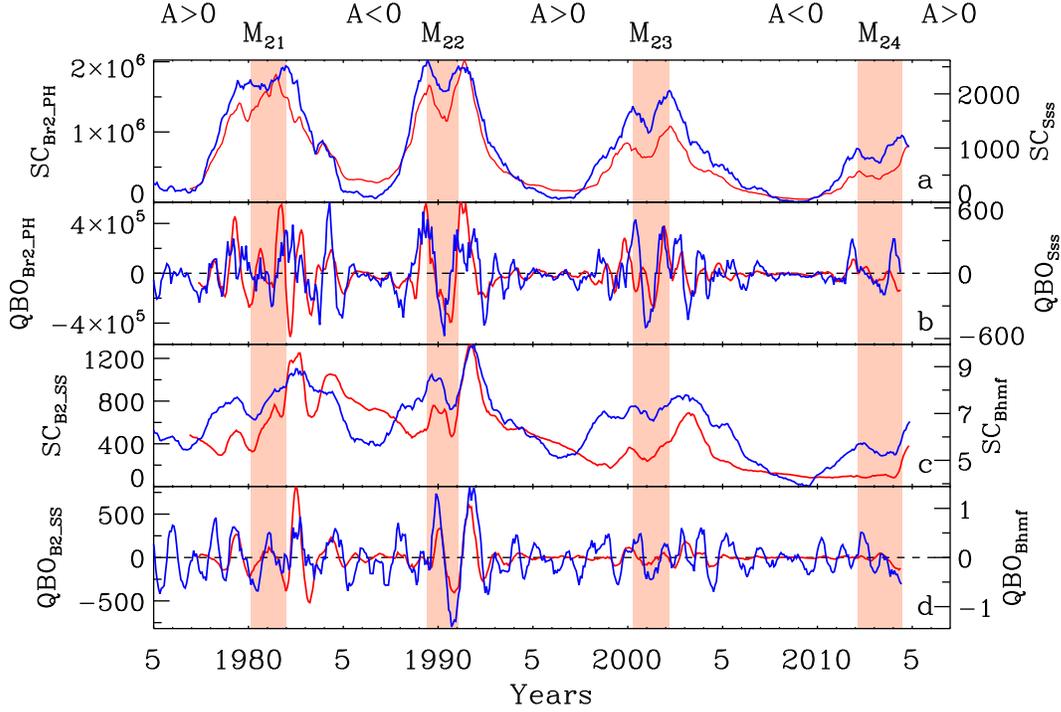}
  \caption{The sunspot cycle and QBO in B2-indices on the photosphere and source surface in 1975-2015 in comparison with the same cycles in the sunspot area and HMF strength.
  The periods of the sunspot maxima are shaded and the HMF polarity $A$ and the moments of the maximum sunspot area are indicated above the panels. In the panels the following variations are shown:
  (a, b) the sunspot cycles and QBOs, respectively, in the radial photospheric SMF energy index
  (red) and in the sunspot area (blue);
  (c,d) the sunspot cycles and QBOs, respectively, in the SMF energy index on the source surface (red) and in the HMF strength near the Earth (blue).
  }
  \label{SC_QBO_B2_PH_SS}
 \end{figure}

In two upper panels (a, b) of Fig. \ref{SC_QBO_B2_PH_SS} the sunspot and QBO cycles in the photospheric radial SMF energy index \verb|Br2_PH| are compared with those of the sunspot area. This comparison supports the conclusion made in  \cite{Bazilevskaya_etal_JoP_CS_inprint_2015,Bazilevskaya_etal_CosRes_inprint_2016}, that these two characteristics change similar both in their long-term and QBO variations. Besides it can be seen that the amplitudes of the QBO are the highest around the sunspot maximum phase (the time period between two Gnevyshev peaks in $B_{ss}$ smoothed with 1-year period, shaded in Fig. \ref{SC_QBO_B2_PH_SS}). Similarly, in two lower panels (c, d) of Fig. \ref{SC_QBO_B2_PH_SS} the sunspot and QBO cycles in the SMF energy index for the source surface, \verb|B2_SS|, are compared with those of the HMF strength $B^{hmf}$. It can be seen from Fig. \ref{SC_QBO_B2_PH_SS} (c) that there is some correlation between the long-term variations of \verb|B2_SS| and $B^{hmf}$, especially during the maximum phase. As to the QBOs in \verb|B2_SS| and $B^{hmf}$ (Fig. \ref{SC_QBO_B2_PH_SS} (d)), they change almost synchronously.

So in the transition from the Sun to the heliosphere the QBO in the SMF energy index demonstrates approximately the same change as that reported in \cite{Bazilevskaya_etal_SSR_186_359_2014} for QBOs in the sunspot activity and the HMF strength. Also in this transition some small shift is observed in the long-term variation of both characteristics.

As it is well known, for $B_r$ in the potential approximation the partial contribution $B_r^l$ from the same degree $l$
changes with $r$ as $\propto r^{-(l+2)}$, so that at the source surface the magnetic field,
which determines the HMF, is influenced mostly by the SMF of the low $l$. It is enticing to suggest that the change in QBOs from the Sun to the heliosphere is due to the same cause. So beside the total SMF energy indices \verb|Br2_PH| and \verb|B2_SS| we constructed the partial SMF energy index \verb|Br2_PH_l|, connected with definite degree $l$:

\begin{equation}
\verb|Br2_PH_l|=\int_0^\pi\int_0^{2\pi}\left.B^l_r\right.^2(r_{ph},\vartheta,\varphi)\sin\vartheta d\vartheta d\varphi\label{Br2_PH_l}
\end{equation}
\noindent and also the similar partial energy index on the source surface, \verb|B2_SS_l|. 
Because of the orthogonality of the spherical functions
$\verb|Br2_PH|=\sum_{l=0}^9 \verb|Br2_PH_l|$ and $\verb|B2_SS|=\sum_{l=0}^9 \verb|B2_SS_l|$.

 \begin{figure}[h]
  \centering
  \includegraphics[width=1.\textwidth]{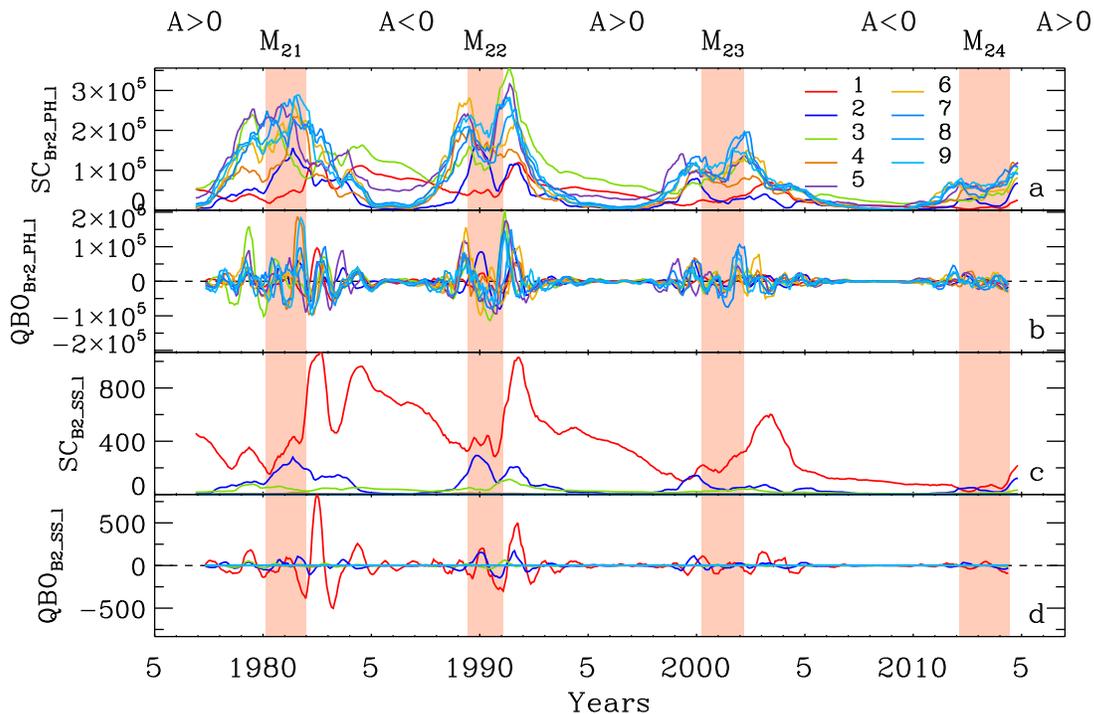}
  \caption{The sunspot cycle and QBO in the partial for each $l$ B2-indices on the photosphere and source surface in 1976-2015. The periods of the sunspot maxima are shaded and the HMF polarity $A$ and the moments of the maximum sunspot area are indicated above the panels. The correspondence between the colors of lines and $l=1\div 9$ is shown in the upper panel. In the panels the following variations are shown:
  (a, b) the sunspot cycle and QBO, respectively, in the radial photospheric SMF partial energy indices
  for different $l=1\div 9$; (c, d) the sunspot cycle and QBO, respectively, in the SMF partial energy indices on the source surface for different $l$.}
  \label{SC_QBO_B2_L_PH_SS}
 \end{figure}

In two upper panels (a, b) of Fig. \ref{SC_QBO_B2_L_PH_SS} the time profiles of the sunspot and QBO cycles in the partial photospheric SMF energy indices \verb|Br2_PH_l| for each $l$ are shown. It could be seen that both for the sunspot and QBO cycles the contributions into the total energy index from the high degrees $l=3,4,7-9$ are significantly greater than from low degrees $l=1,2$ and the time profile of the partial energy index with $l=1$ is somewhat lags behind the more powerful partial indices. In two lower panels of Fig. \ref{SC_QBO_B2_L_PH_SS} the sunspot and QBO cycles in the partial SMF energy indices \verb|B2_SS_l| for each $l$  are shown. It is clearly seen that on the source surface and hence in the HMF for both long-term and QBO cycles the low-$l$ partial indices are the most important. So in the first approximation our hypothesis is justified that the change in the sunspot and QBO cycles in the transition from the Sun to the heliosphere is due to 1) the different magnitude and time behavior of the large-scale (low $l$) and small-scale (high $l$) photospheric SMF and 2) the stronger attenuation of the SMF with higher $l$ in this transition. The first of these facts and how it correlates with the sunspot distribution still should be thought over. The conclusion in \cite{Bazilevskaya_etal_SSR_186_359_2014} that the 11-year variation in contrast with QBO does not changes its phase during this transition is probably due to the fact that the observed lag of the time profiles in the heliosphere with respect to those on the Sun (approximately similar for both variations) for 11-y cycle is much smaller than its period.

Note that the partial SMF energy indices can be constructed not only for the whole spheres but for different ranges of latitude, e.g., for different hemispheres or the royal zones (the latitude ranges with the sunspots) \cite{Krainev_etal_Pulkovo_Conf_121_1999}. We are planning to use them when looking for the explanation of, e.g., the different QBOs in the sunspot activity in the N- and S-hemispheres, reported in  \cite{Bazilevskaya_etal_SSR_186_359_2014}.

\section{On the causes and mechanisms of the QBO in GCR intensity}
It was suggested in \cite{Bazilevskaya_etal_SSR_186_359_2014} that the probable cause of the $QBO_J$ is the opposite in sign variation in the HMF strength, $QBO_{Bhmf}$, and it is indicative of the diffusion as a main mechanism of the $QBO_J$. Besides, the intermittent QBO in the heliospheric current sheet (HCS) quasi-tilt, $QBO_{\alpha_{qt}}$, \cite{Bazilevskaya_etal_SSR_186_359_2014} can play some role in the difference between the $QBO_J$ for the periods of the opposite HMF polarity. Here we note that change of $B^{hmf}$ influences not only the diffusion coefficient ($K_{diff}\propto 1/B^{hmf}$ in many models of the GCR intensity modulation), but also the magnetic drift velocity (also $V_{drift}\propto 1/B^{hmf}$). So the same $QBO_{Bhmf}$ in different periods could not only result in the same $QBO_J$, but also along with $QBO_{\alpha_{qt}}$ could give rise to different $QBO_J$ for the $A>0$ and $A<0$ periods. Moreover, as the contribution of the magnetic drift into the 11-year variation of the GCR intensity could be significant (see \cite{Krainev_Kota_Potgieter_icrc2015-198} and references therein), the same $QBO_J$ from the same $QBO_{Bhmf}$ could be the result not only of the diffusion but of the magnetic drift as well.

 \begin{figure}[h]
  \centering
  \includegraphics[width=0.75\textwidth]{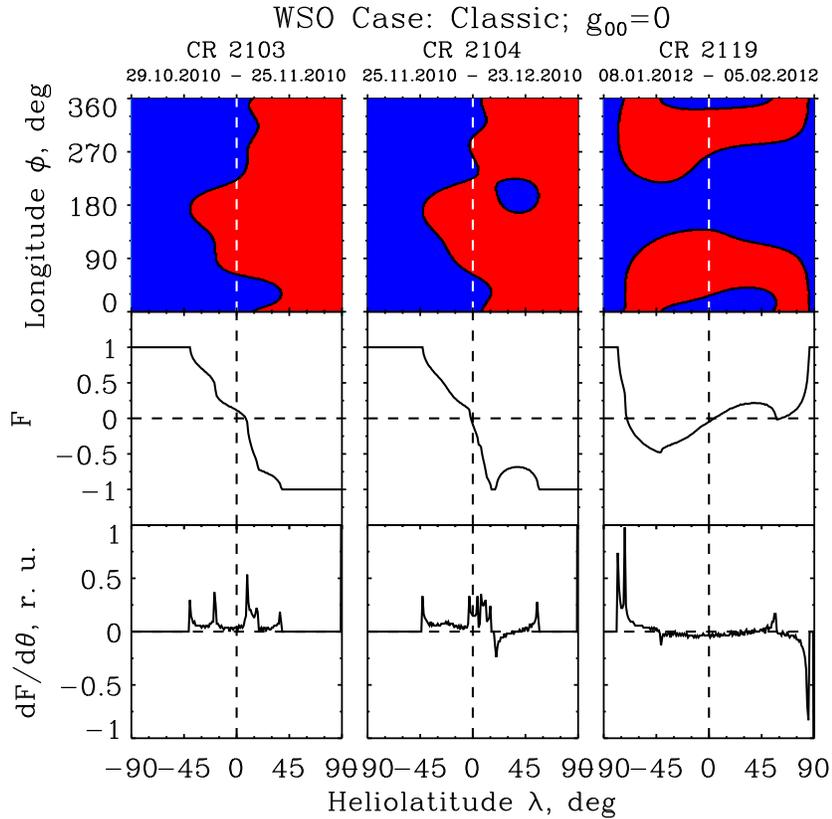}
  \caption{The map of the HMF polarity ${\cal F}$, the longitude averaged HMF polarity $F$ and its the polar angle derivative $dF/d\vartheta$ for three types of the HMF polarity distribution. The panels in the upper row are the HMF polarity (${{\cal F}}$) distributions for the "dipole" type (left), the "transition dipole" (middle) and  "inversion" (right) types. The color (red for negative and blue for positive) stands for the HMF polarity, the black lines between being the HCSs. In the panels in the middle and lower rows $F$ and $dF/d\vartheta$, respectively, are shown for the corresponding HMF polarity distributions of the upper row.}
  \label{Maps_F_dFdt_3_Types}
 \end{figure}

Our second note on the mechanisms of the QBO in GCR intensity concerns the description of the magnetic drift during the maximum phase of the sunspot cycle. For the phase of the low sunspot activity with the ``dipole'' type of the HMF polarity distribution (the single global HCS connecting all longitudes; see the HMF polarity classification in \cite{Krainev_etal_icrc2015-437}), the GCR intensity can be calculated using the transport equation with the usual magnetic drift velocity terms, e.g., utilizing the tilted-HCS model with a tilt $\alpha_t$ as a parameter, and getting $\alpha_t$ as the quasi-tilt $\alpha_{qt}$ from \cite{WSO_Site}. However, it is not so simple to get these terms for the high sunspot activity phase with the ``transition dipole'' and  ``inversion'' types of the HMF polarity distributions, when there are several HCSs and the use of the formally defined quasi-tilt is questionable.

For the time being we use the following procedure.
As in \cite{Kalinin_Krainev_ECRS21_222-225_2009} the regular 3D HMF can be represented as $\vec{\cal B}(r,\vartheta,\varphi)={\cal F}(r,\vartheta,\varphi)\vec{\cal B}^m(r,\vartheta,\varphi)$,
where $\vec{\cal B}^m$ is the unipolar
(or ``monopolar'', $B^{hmf}_r>0$ everywhere) magnetic field and the HMF polarity ${\cal F}$ is a scalar function equal to $+1$ in the positive and $-1$ in negative sectors,
changing on the HCS surface ${{\cal F}} (r,\vartheta,\varphi)=0$.
Then the 3D particle drift velocity is
${\vec{{\cal V}}}^d=pv/3q\left[{\bf{\nabla}}\times(\vec{{\cal B}}/{{\cal B}}^2)\right]$, \cite{Rossi_Olbert_1970}, where $v$ and $q$ are the particle speed and
charge, respectively. Then one can decompose the drift velocity into the regular and current sheet velocities:
\begin{eqnarray}
{\vec{{\cal V}}}^{d,reg}=pv/3q{{\cal F}}\left[{\bf{\nabla}}\times(\vec{{\cal B}}^m/{{\cal B}}^2)\right]\label{Vreg}\\
{\vec{{\cal V}}}^{d,cs}=pv/3q\left[{\bf{\nabla}}{{\cal F}}\times(\vec{{\cal B}}^m/{{\cal B}}^2)\right].\label{Vcs}
\end{eqnarray}

So to get the magnetic drift velocities for any type of the HMF polarity distribution in 3D case one needs only ${\cal F}$ and ${\bf{\nabla}}{\cal F}$ and in 2D case the averaged over the longitude $F$ and $dF/d\vartheta$. All of these quantities (${{\cal F}},\nabla{\cal F},F,dF/d\vartheta$) can be calculated numerically for any calculated HMF polarity distribution, including the "transition dipole" and  "inversion" types. In Fig. \ref{Maps_F_dFdt_3_Types} along with the maps of ${{\cal F}}(\vartheta,\varphi)$ on the source surface, the averaged over the longitude $F$ and $dF/d\vartheta$ are shown as functions of latitude for three types of the HMF polarity distribution.
It can be seen that for the ``dipole'' type (the left column of Fig. \ref{Maps_F_dFdt_3_Types}) both $F$ (and hence the regular drift velocity (\ref{Vreg})) and $dF/d\vartheta$ (and hence the HCS-drift velocity (\ref{Vcs})) have the usual (although somewhat irregular) appearance similar to the case of simple tilted-HCS. The addition to the global HCS of one more local HCS (the middle column of Fig. \ref{Maps_F_dFdt_3_Types}) adds two features of opposite sign to $dF/d\vartheta$ (and hence two streams of opposite direction to HCS-drift velocity) at the latitudinal boundaries of the additional HCS. At last, the ``inversion'' type of the HMF polarity distribution (the right column of Fig. \ref{Maps_F_dFdt_3_Types}) changes drastically both  $F$ and $dF/d\vartheta$, so that the HMF polarity in the polar regions are of the same signs and in these regions there are two strong HCS-streams of opposite direction. In 3D case there are also strong meridional flows along the HCSs and so the overall picture of the regular and HCS drifts looks rather systematic and intriguing. Note, however, that the HCS-drift velocity (\ref{Vcs}) does not take into account the smearing of the current sheet drift due to the finite larmor radius of the GCR particles \cite{Burger_etal_ApSS_116_107_1985}. In \cite{Krainev_Kalinin_33ICRC_317_2013} we suggested that the fundamental difference between the global and nonglobal HCS lies in the fact that the sign of the radial component of the current sheet drift changes as the particle moves along the  nonglobal HCS, so that the connection between the inner and outer heliosphere is blocked.

So with respect to the HMF polarity distribution the conditions in the heliosphere are quite different around the sunspot maximum and during the periods of low solar activity. If we take into account the presence during high solar activity of the global merged interactive regions serving as barriers for the GCR propagation \cite{LeRoux_Potgieter_ApJ_442_847_1995}, the suggestion made in \cite{Bazilevskaya_etal_SSR_186_359_2014} that both step-like changes of the GCR intensity and Gnevyshev Gap effect could be viewed as the manifestations of the QBO in the GCR intensity, that is, could have the same nature, looks questionable.

\section{Acknowledgments}
The authors thank for help the Russian Foundation for Basic Research (grants 13-02-00585, 14-02-00905)
and the Program of the Presidium of the Russian Academy of Sciences ``Fundamental Properties of Matter and Astrophysics''.


\begin{thebibliography}{99}

\bibitem{Bazilevskaya_etal_SolPhys_197_157_2000}
G.A. Bazilevskaya et al., \emph{Structure of the maximum phase of the solar cycles 21 and 22}, \emph{Solar Phys.} {\bf 197} (2000) 157-174.

\bibitem{Bazilevskaya_etal_SSR_186_359_2014}
G. Bazilevskaya et al., \emph{A combined analysis of the observational aspects of the quasi-biennial oscillation in solar magnetic activity}, \emph{Space Sci. Rev.} {\bf 186} (2014) 359-386.

\bibitem{Bazilevskaya_etal_JoP_CS_inprint_2015}
G. Bazilevskaya et al., \emph{Correlation of the quasi-biennial oscillations in
galactic cosmic rays and in the solar activity indices}, \emph{J. Physics: Conf. Series} in print (2015).

\bibitem{Bazilevskaya_etal_CosRes_inprint_2016}
G. Bazilevskaya et al., \emph{On the correlation between the quasi-biennial oscillations in
 the solar activity, interplanetary magnetic field and cosmic rays}, \emph{Cosmic Research} in print (2015).

\bibitem{Burger_etal_ApSS_116_107_1985}
R.A. Burger, H. Moraal, G.M. Webb, \emph{Drift theory of charged particles in electric and magnetic fields},
\emph{Astrophys. Space Sci.} {\bf 116} (1985) 107.

\bibitem{Kalinin_Krainev_ECRS21_222-225_2009}
M.S. Kalinin, M.B. Krainev, \emph{On the wavy heliospheric current sheet in the 2D transport equation for the galactic cosmic rays}, in proceedings of \emph{21st European Cosmic Ray Symposium} (2009) 222-225.

\bibitem{Krainev_etal_Pulkovo_Conf_121_1999}
M.B. Krainev, G.A. Bazilevskaya, V.S. Makhmutov, \emph{On the large scale solar magnetic fields and the double peak structure of solar activity during the maximum phase of solar cycle}, in proceedings of \emph{Internat. Conf. "Large-scale structure of solar
activity: achievements and perspectives"} (1999) 121, Pulkovo, St.-Petersburg.

\bibitem{Krainev_Webber_IAUSymp_223_81_2004}
M.B. Krainev, W.R. Webber, \emph{The Solar Cycle in the heliospheric parameters and galactic cosmic ray intensity}, in proceedings of \emph{IAU Symp. No. 223.} A.V. Stepanov etal., eds. (2004) 81-84.

\bibitem{Krainev_Kalinin_33ICRC_317_2013}
M.B. Krainev, M.S. Kalinin, \emph{On the GCR intensity and the inversion of the heliospheric magnetic field during the periods of the high solar activity}, in proceedings of \emph{33rd International Cosmic Ray Conference} ID-0317 (2013), [{\tt astro-ph.SR/411.7532}].

\bibitem{Krainev_etal_icrc2015-437}
M. Krainev et al., \emph{GCR intensity during the sunspot maximum phase
and the inversion of the heliospheric magnetic field}, This conference ID-437 (2015)

\bibitem{Krainev_Kota_Potgieter_icrc2015-198}
M.B. Krainev, J. K\'ota, M.S. Potgieter, \emph{On the causes and mechanisms of the long-term variations in the GCR characteristics},
This conference ID-198 (2015)

\bibitem{LeRoux_Potgieter_ApJ_442_847_1995}
J.A. le Roux and M.S. Potgieter, \emph{The simulation of complete 11 and 22 year modulation cycles for cosmic rays in the heliosphere using a drift model with global interaction regions}, \emph{Astrophysical Journal} {\bf 442} (1995) 847-851.

\bibitem{Obridko_Shelting_SP_137_167_1992}
V.N. Obridko, B.D. Shelting, \emph{Cyclic variation of the global magnetic field indices}, \emph{Solar Physics} {\bf 137} (1992) 167.

\bibitem{Rossi_Olbert_1970}
B. Rossi, S. Olbert, \emph{Introduction to the Physics of Space}, McGraw-Hill (1970) p.111.

\bibitem{OMNI_Site}
\verb|ftp://omniweb.gsfc.nasa.gov/pub/data/omni|

\bibitem{Sss_Site}
\verb|http://solarscience.msfc.nasa.gov/greenwch.shtml|

\bibitem{WSO_Site}
\verb|http://wso.stanford.edu/|

\end{thebibliography}
\end{document}